\documentclass[aps,prb,twocolumn,prb]{revtex4-1}
\usepackage{amsmath}
\usepackage{natbib}
\usepackage[pdftex]{graphicx}
\usepackage{mathtools}

\usepackage[usenames]{color}

\begin{document}

\date{\today}

\author{Brankica Jankovic, Olga Bozovic, Peter Hamm$^*$}
\affiliation{Department of Chemistry, University of Zurich, Winterthurerstr.\ 190, CH-8057 Z\"urich, Switzerland\\$^*$peter.hamm@chem.uzh.ch}

\title {Intrinsic Dynamics of Protein-Peptide Unbinding}

\begin{abstract}
\textbf{Abstract}: The dynamics of peptide-protein binding and unbinding of a variant of the RNase~S system has been investigated. To initiate the process, a photoswitchable azobenzene moiety has been covalently linked to the S-peptide, thereby switching its binding affinity to the S-protein. Transient fluorescence quenching was measured with the help of a time-resolved fluorometer, which has been specifically designed for these experiments and is based on inexpensive LED's and laser diodes only. One mutant shows on-off behaviour with no specific binding detectable in one of the states of the photoswitch. Unbinding is faster by at least two orders of magnitudes, as compared to other variants of the RNase~S system. It is concluded that unbinding is essentially barrier-less in that case, revealing the intrinsic dynamics of the unbinding event, which occurs on a few 100~$\mu$s timescale in a strongly stretched-exponential manner.
\end{abstract}

\maketitle

\section{Introduction}

Proteins employ sophisticated binding mechanisms during their interplay with other physiological partners to fulfill crucial biological processes. Starting from the recognition of small rigid molecules by proteins  all the way to complex rearrangements of protein-protein interactions, many different models of protein binding have been suggested.\cite{vogt2012, gianni2014, hammes2009, paul2016} The important transient conformational changes associated with binding can be hidden in equilibrium structures, and capturing them is the only way to provide comprehensive mechanistic insights. One example of a challenging question related to protein interactions is related to intrinsically disordered proteins. These proteins interact by coupling binding and folding and a lot of effort has been directed towards understanding the temporal ordering of the underlying events.\cite{wright2009} Therefore,  clarifying the detailed binding mechanisms of protein interactions necessitates a kinetic perspective.\cite{chakrabarti2016l}

NMR-based techniques revealed invaluable insights into the complicated binding mechanisms of proteins, such as the ``fly-casting'' interaction mechanism of the intrinsically disordered pKID transcription factor with the KIX domain, i.e., a hydrophobic loose encounter complex is followed by the second (folding) phase.\cite{sugase2007} However, approaches that allow for a direct observation of events related to binding in a time-resolved manner are very scarce. They are commonly based on the stopped-flow methods,\cite{shammas2012, rogers2013, rogers2014} where rapid mixing of two interacting species is used as a trigger to initiate binding (or unbinding in competition experiments).\cite{shammas2016, gianni2016} These methods are limited by the mixing time, leaving the accessible time window in the millisecond regime.\cite{gianni2016} On the other hand, there are ultrafast laser-based approaches that rely on a temperature jump as a trigger for conformational changes, and fluorescence as a way of detection.\cite{dosnon2015}  The different phases in the binding process of a yeast protease and its intrinsically disordered inhibitor have been resolved in this way.\cite{narayanan2008} However, temperature jumps, which are typically small, are limited by the size of the perturbation that can be induced. Single molecule fluorescence revealed the electrostatically driven encounter complex formation, followed by folding into the final 3D conformation.\cite{sturzenegger2018} A recent MD simulation revealed an atomistic support for the induced fit binding mechanism of an intrinsically disordered system \cite{robustelli2020}. Key contacts between the disordered peptide and the protein formed before or in parallel with the secondary structure formation. However, MD simulations are limited to very fast binders, and it is still very difficult to approach the relevant range of milliseconds.

The emerging strategy of designing photoswitchable proteins and peptides proved fruitful for diverse studies,\cite{Beharry2011} where a precise control of certain aspects of protein structure and/or function is necessary to control for instance protein folding,\cite{spo02,woolley05,rehm05,aemissegger05,schrader07,Ihalainen2008,Lorenz2016} allosteric communication,\cite{buchli13,stock2018, bozovic2020b, bozovic2020a} or biological activity.\cite{hoorens2018, schierling2010, brechun2017, zhang2010} Here we employ the previously designed photoswitchable RNase S to explore the kinetics and dynamics of (un)binding of this non-covalent complex.\cite{jankovic2019} This model system has been previously used to study the mechanism of coupled binding and folding,\cite{goldberg1999, bachmann2011} as the S-peptide fragment is unfolded in isolation, while it adopts the helical structure once bound to the S-protein part.\cite{Luitz2017, schreier1977, goldberg1999, richards1959, bachmann2011}

\begin{figure}[t]
	\centering
	\includegraphics[width=0.45\textwidth]{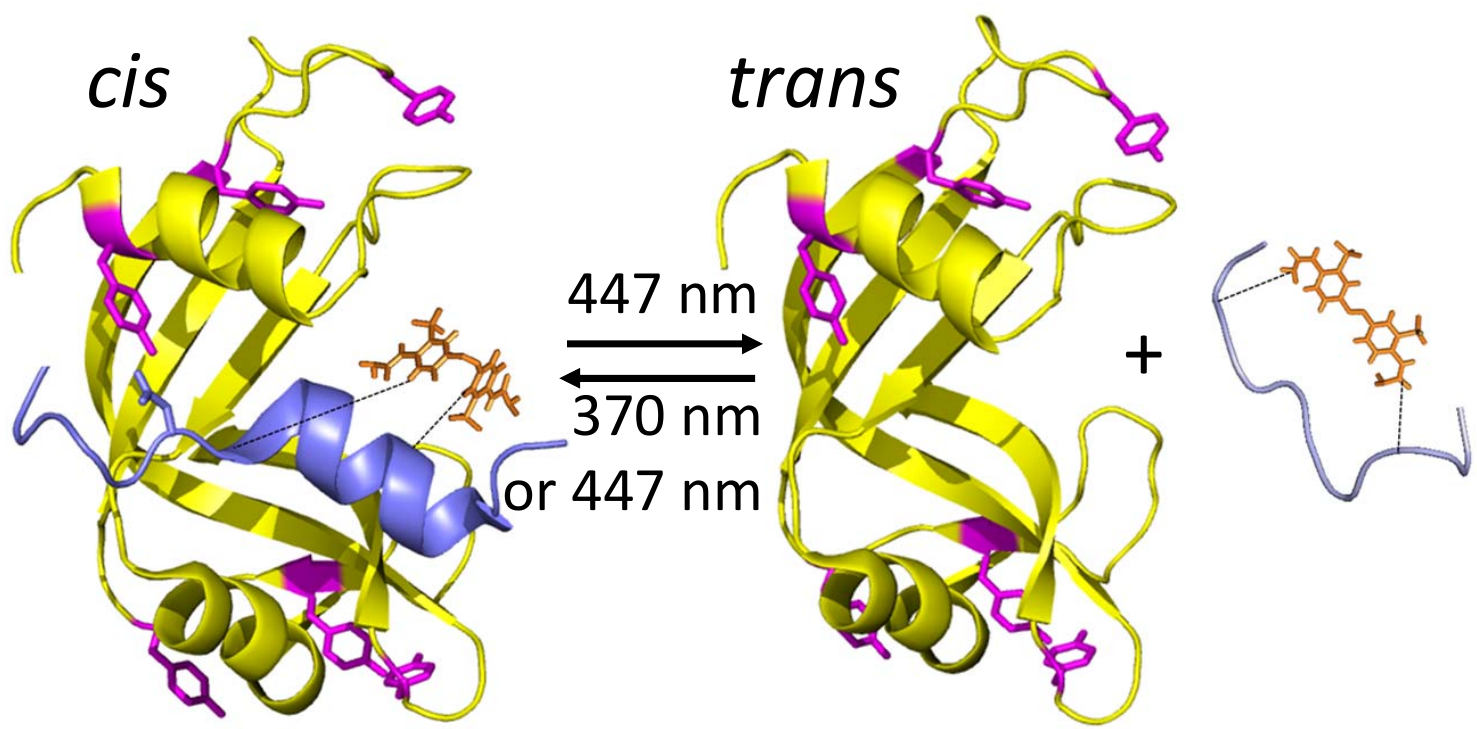}
	\caption{Molecular construct.\cite{jankovic2019} The S-protein (yellow) with highlighted tyrosine residues (magenta) binds the S-peptide (blue) in the \textit{cis}-state of the azobenzene moiety (orange), while it unbinds in the \textit{trans}-state. The picture was adapted from pdb-entry 2e3w.\cite{Boerema2008}} \label{figStruct}
\end{figure}

Our molecular construct is illustrated in Fig.~\ref{figStruct}.\cite{jankovic2019} The azobenzene moiety (orange) is covalently linked to the S-peptide (blue) via two cysteines (see Methods for details). By choosing the distance between these anchoring points, the $\alpha$-helicity of the S-peptide in the two states (\textit{cis} or \textit{trans}) of the photoswitch is either stabilized or destabilized,\cite{flint02} which in turn determines its binding affinity to the S-protein (yellow).\cite{jankovic2019} One can selectively switch between both states with light of the proper wavelength. In Ref.~\onlinecite{jankovic2019}, we designed five different mutants with varying anchoring points, and in one case with an additional mutation. The binding affinities of the S-peptide to the S-protein in the \textit{cis} and the \textit{trans}-states of the photoswitch have been measured by a combination of ITC, CD spectroscopy and intrinsic tyrosine fluorescence quenching. As anticipated by our design, the binding affinity is larger in the \textit{cis}-state for all mutants we investigated. However, the values for the binding affinities, and in particular the factors by which the binding affinity changes upon switching, vary significantly. S-pep(6,13), with the photoswitch linked at positions 6 and 13, sticks out in this regard, as it binds with reasonable affinity in the \textit{cis}-state, but no specific binding could be detected by CD spectroscopy in the \textit{trans}-state (fluorescence quenching indicated some degree of unspecific binding). This mutant will be the focus of the present kinetic study, as it approaches the ``speed limit'' of ligand unbinding and thus reveals its intrinsic dynamics, analogous to the concept of downhill protein folding.\cite{sabelko99,Yang03}  As a control, we will also consider S-pep(6,10), which has a large (20 fold) change in binding affinity, but stays specifically bound to the protein in both states.\cite{jankovic2019}

In either case, the photoswitchable S-peptide does not have any fluorophore, while the S-protein has six tyrosine residues, one of which is located in the binding groove (Fig.~\ref{figStruct}). The amount of fluorescence will be sensitive to peptide binding, as the peptide (presumably mostly its azobenzene moiety) quenches the tyrosine fluorescence; this is the effect that enables one to determine binding affinities from the concentration dependent fluorescence yield.\cite{jankovic2019} Here, we measure fluorescence in a transient manner in order to follow the kinetics and/or dynamics of ligand binding and unbinding.

\section{Methods}
\begin{figure}[t]
	\centering
	\begin{center}
		\includegraphics[width=.45\textwidth]{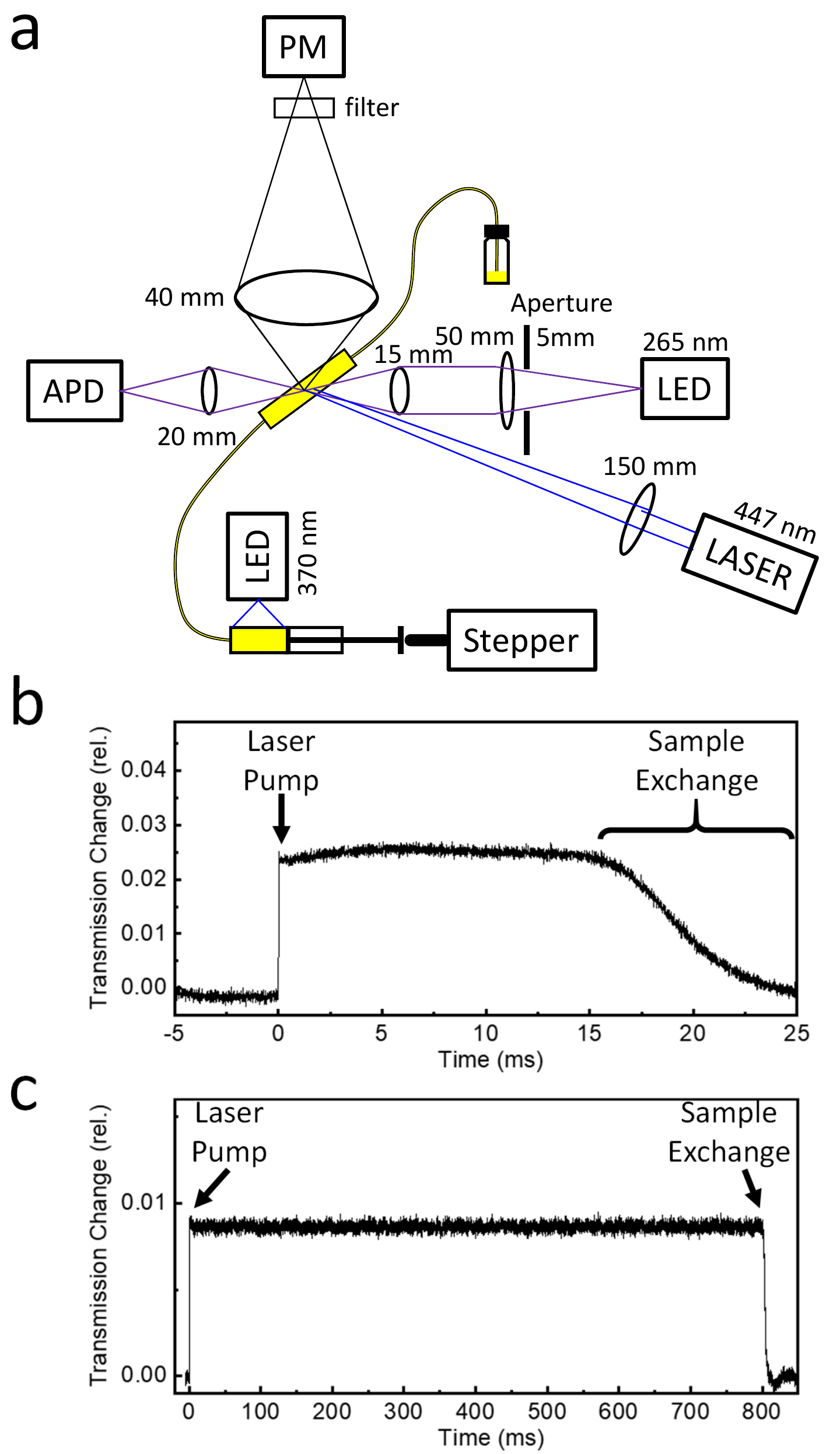}
		\caption{Time-resolved fluorometer. (a) Experimental setup used in this study; the various components are discussed in the text. Test experiments demonstrating the performance of the system are shown in panel (b) for a high frequency (33 Hz) and in panel (c) for a low frequency (1.1 Hz) of the syringe pushes.}\label{FigSetup}
	\end{center}
\end{figure}

\subsection{Experimental Setup}
Fig.~\ref{FigSetup}a shows the experimental setup, which has been specifically designed for this study. A UV-LED at 265~nm with active area 1~mm$^2$ (M265D2, Thorlabs) was used to excite the fluorescence of the tyrosine residues of the protein sample. It was operated by a pulsed laser diode driver (LDP-V~10-10, PicoLAS), producing 15~ns pulses at a repetition rate of 200~kHz. The light was collected with a 50~mm lens, spatially filtered with an aperture with dia=5~mm, and focused with a 15~mm lens into the sample with a spot size of $\approx$300~$\mu$m. We estimated that the time-averaged power in the sample was $\approx1~\mu$W. A low power was anticipated to minimize the number of molecules that photo-isomerize induced by that measurement light (we estimated that it takes about a minute until every molecule in the measurement volume would have seen a 265~nm photon). The transmitted light was measured by an avalanche photo diode (APD, APD120A2/M, Thorlabs). The fluorescence light was collected in a 90$^\circ$ geometry by an large-aperture aspherical lens ($f$=40~mm, dia 50~mm), spectrally filtered with an interference filter transmitting $\approx$ 300-360~nm (XRR0340, Asahi Spectra), and imaged onto a photomultiplier (PM, PMA~175-N-M, PicoQuant). The signals from the APD and PM have been digitized in a home-built 16 bit ADC (similar to the one described in Ref.~\onlinecite{Farrell2020}), and transferred to a computer for data processing.

Pump pulses at 447~nm were generated with a GaN laser diode (PLPT9~450D\_E, Osram Opto Semiconductors), operated by another pulsed laser diode driver (LDP-V~10-10, PicoLAS) to produce pulses of 2~$\mu$s length at typical repetition frequencies of 1-33~Hz. While the maximum power of laser diode is specified at 3.5~W in cw-operation, we found that one can go up to 10~W in pulsed operation, revealing 20~$\mu$J of pulse energy in the 2~$\mu$s long pulses. The laser diode beam was pre-collimated (LTN330-C, Thorlabs), its elliptical shape corrected with two cylindrical lenses (50~mm and 150~mm), and then focused into the sample with a 150~mm lens, roughly matching the diameter of the probe light.

The intrinsic time resolution of the setup is 5~$\mu$s, determined by the repetition rate (200~kHz) of the UV-LED used to excite the fluorescence of the tyrosine residues. The signal-to-noise is inversely proportional to the square root of the number of detected fluorescence photons, which is shot-noise limited. As shot-noise is uncorrelated (i.e., white noise), signal-to-noise can be improved by time-filtering data with a Gaussian function, at the expense of time resolution. This effectively increases the number of detected photons, and signal-to-noise is inversely proportional to the square root of the effective time resolution after time-filtering.

The experiment required the exchange of sample between subsequent excitations from the 447~nm laser, which was achieved with a pulsed syringe pump pushed by a stepper motor (DRV014, Thorlabs). The stepper motor controller (KST101, Thorlabs) has an  external trigger input, moving the stepper motor at desired time points with steps whose size can be pre-programmed. The syringe was connected to the sample cuvette\cite{Bredenbeck2003a} via rigid Teflon tubings. The channel in the sample cuvette was about 1~mm wide and 200~$\mu$m thick. The ratio of the dimension of the syringe vs that of the channel in the sample cuvette translated the 4~$\mu$m  steps of the stepper motor into the desired $\approx 300~\mu$m steps of the excited spot in the sample cuvette. The stepper motor controller had to be programmed in Labview, allowing one to overwrite the default settings for the maximal velocity and acceleration, which was needed for stepping frequencies up to 33~Hz. A total sample volume of about 1~ml was needed. The timings of all components have been controlled with a programable delay generator (T560, Highland Technology).

\begin{figure}[t]
	\centering
	\begin{center}
		\includegraphics[width=.45\textwidth]{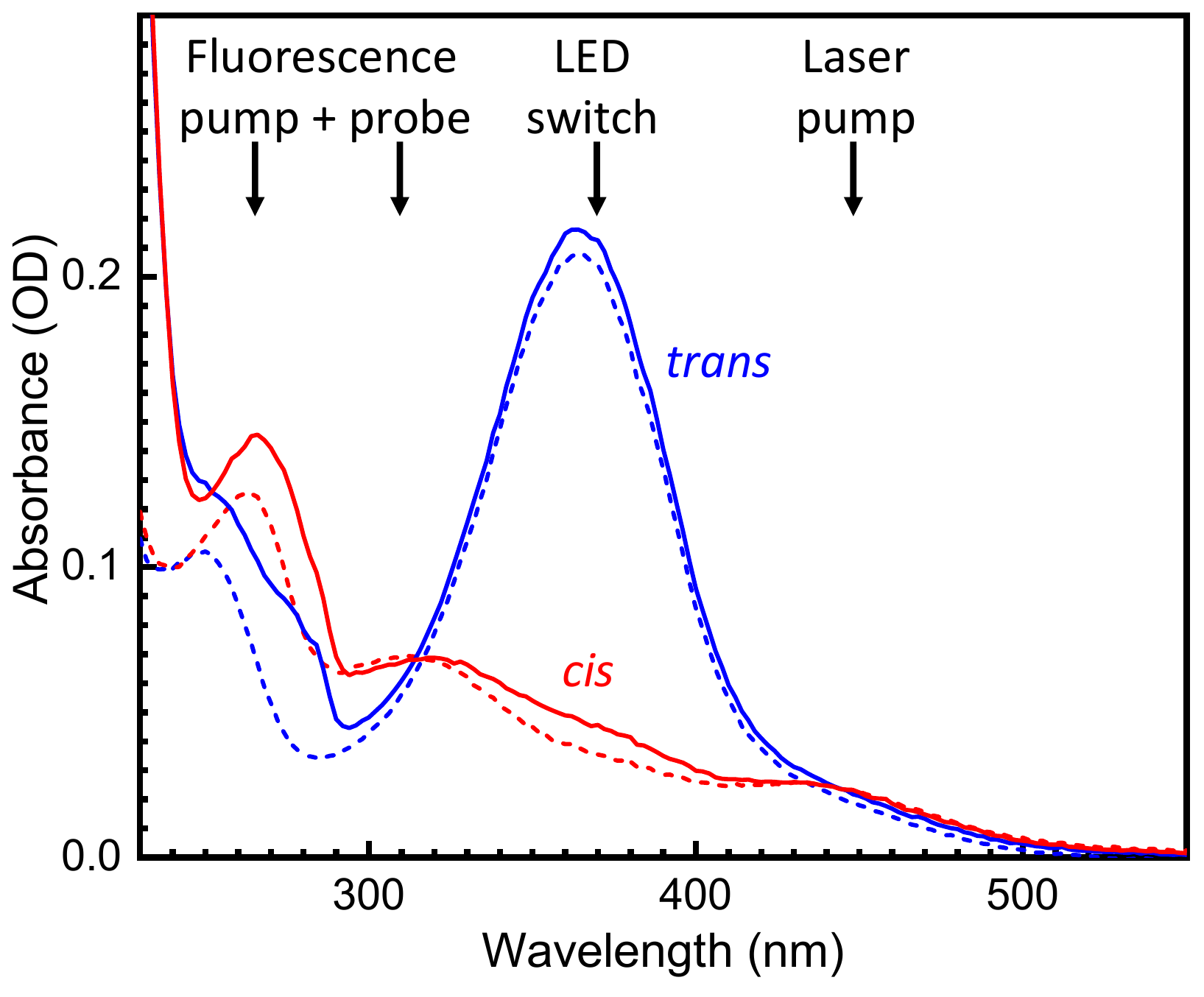}
		\caption{Absorption spectra. Absorption spectra exemplified for S-pep(6,13) (dashed lines) and S-pep(6,13)+S-protein (solid lines) in their \textit{cis} (red) and \textit{trans} (blue) states. The arrows indicate the wavelengths of the various light sources used in the experimental setup, as well as that of the fluorescence emission.}\label{FigSpec}
	\end{center}
\end{figure}

Figs.~\ref{FigSetup}b,c demonstrate the performance of the pulsed syringe pump, using only the azobenzene-photoswitch (without protein, the part colored in orange in Fig.~\ref{figStruct}) as test sample. To this end, the sample was first prepared in its \textit{cis}-state  with an excess of 370~nm light from a cw-LED, which illuminated the fused silica syringe. The $trans$-absorption is significantly larger at this wavelength (see Fig.~\ref{FigSpec}), thereby shifting the photo-equilibrium to the $cis$-state with typically 85\%. The 447~nm laser pulse then induced a \textit{cis}-to-\textit{trans} isomerization at time-zero with a quatum yield of $\approx$60\%.\cite{borisenko05}

Fig.~\ref{FigSetup}b shows that photo-isomerization is instantaneous on the timescale of this experiment, revealing a step-like increase in transmission of the 265~nm light from the UV-LED at time-zero, since the absorption of the photoswitch changes upon photo-isomerization (see Fig.~\ref{FigSpec}). The signal stays roughly constant for $\approx$15~ms after laser excitation, which will be the usable time window for measuring transient fluorescence, followed by a period of $\approx$10~ms for sample exchange upon pushing the syringe pump. The overall data are periodic, and we shifted the subsequent 5~ms, which are used to determine an offset, to negative times in Fig.~\ref{FigSetup}b. The dead time of the experiment is thus $\approx$15~ms. When a lower repetition rate is chosen, the dead time remains the same, while the usable time window increases accordingly, see Fig.~\ref{FigSetup}c (some of the protein samples were ``sticky'', hampering a smooth motion of the piston in the syringe, in which case it was necessary to increase the time between the syringe pushes and the 447~nm laser pulses to 100-250~ms). This plot also shows that the sample does not move and/or diffuse on a 1~s timescale between the syringe pushes. For a quick and complete exchange of the sample, it turned out to be very critical that absolutely no bubbles were present in the syringe, teflon tubings or the sample cuvette, and that the spatial overlap between the 265~nm probe light and the 447~nm laser pulses was carefully aligned.

The opposite \textit{trans}-to-\textit{cis} switching direction (ligand binding) could also be measured with this setup. To that end the sample was kept in the dark, in which case it would eventually relax into the lower-energy \textit{trans}-state. Illumination with the 447~nm laser then induced \textit{trans}-to-\textit{cis} isomerization, since in essence no \textit{cis}-peptides were present and since the \textit{trans}-state also absorbs at this wavelength (see Fig.~\ref{FigSpec}). The quantum yield is however significantly lower in this case.\cite{borisenko05} Photo-isomerisation competes with thermal \textit{cis}-to-\textit{trans} back relaxation, but since the time-averaged power of the 447~nm laser is very small (40~$\mu$W at 2~Hz), the photo-equilibrium of the sample as a whole will almost exclusively be on the \textit{trans}-side.

\subsection{Sample Preparation}

The S-protein was prepared by cleaving the commercial ribonuclease A from bovine pancreas (Sigma-Aldrich) with subtilisin (Sigma-Aldrich), as described in Refs.~\onlinecite{richards1959,jankovic2019} (with small modifications).  To limit the proteolysis to a single peptide bond (between residues 20 and 21), we performed the cleavage reaction on ice overnight. The reaction was stopped by adjusting the pH value to 2. The S-protein was purified by C5 reverse-phase chromatography.

Photoswitchable peptides were prepared by crosslinking the cysteine-containing peptides with the water-soluble azobenzene-based photoswitch.\cite{zhang03} The peptides were first synthesized by standard Fmoc-based solid-phase peptide-synthesis using a Liberty 1 peptide synthesizer (CEM Corporation, Matthews, NC, USA). All amino acids were purchased from Novabiochem (La Jolla, CA, USA). The photoswitch (3,3'-bis(sulfonato)-4,4' bis(chloroacetamido)azobenzene) was added to a peptide reduced by tris(2-carboxyethyl)phosphine (TCEP) in 5x molar excess and incubated overnight. The linked peptides were purified by C18 reverse-phase chromatography.

The purity of all protein and peptide samples was analyzed by mass spectrometry. Concentrations were determined by amino acid analysis. All solutions were prepared in 50 mM sodium phosphate buffer pH 7.0.\\

\subsection{Model}

For a quantitative determination of the on- and off-rate constants, we considered the following coupled equilibria, in which two molecular species, the S-peptide in its \textit{cis} and \textit{trans}-states, compete for the same binding site on the S-protein $P$:
\begin{align}
PL_{cis}&\xrightleftharpoons[k_{on,cis}]{k_{off,cis}} P+L_{cis} \nonumber\\
PL_{trans}&\xrightleftharpoons[k_{on,trans}]{k_{off,trans}} P+L_{trans} \label{eqEqcouled}
\end{align}
The corresponding differential equations were solved numerically with the help of Mathematica. For the initial conditions, we first determined the equilibrium conditions in either the \textit{cis} or the \textit{trans}-state, assuming the S-peptide is 100\% in this state, and then switched 10\% or 5\% of the molecules for the \textit{cis}-to-\textit{trans} or \textit{trans}-to-\textit{cis} isomerization, respectively (accounting for the smaller isomerization quantum yields of the latter).
Despite the fact that the solutions of these differential equations are not strictly exponential, the deviation from exponential is very small and the data from the model
were fit to single-exponential functions.
The model is too simple to expect a quantitative fit of the experimental data; for example it ignores the possibility of unspecific binding, while we have evidence from comparing CD and fluorescence binding curves that unspecific binding does exist to a certain extent.\cite{jankovic2019} We therefore concentrated on the time constants; in the case of \textit{trans}-to-\textit{cis} switching only on those at lower S-protein concentration, when the effect of unspecific binding is expected to be less. Binding affinities known from Ref.~\onlinecite{jankovic2019} were taken over, and the other parameters of the model ($k_{on,cis}$ and $k_{on,trans}$ for S-pep(6,10), and $k_{on}$ and $K_{d,trans}$ for S-pep(6,13)) were varied until similar time constants as in experiment were obtained.

\section{Results}

\begin{figure*}[t]
	\centering
	\begin{center}
		\includegraphics[width=.8\textwidth]{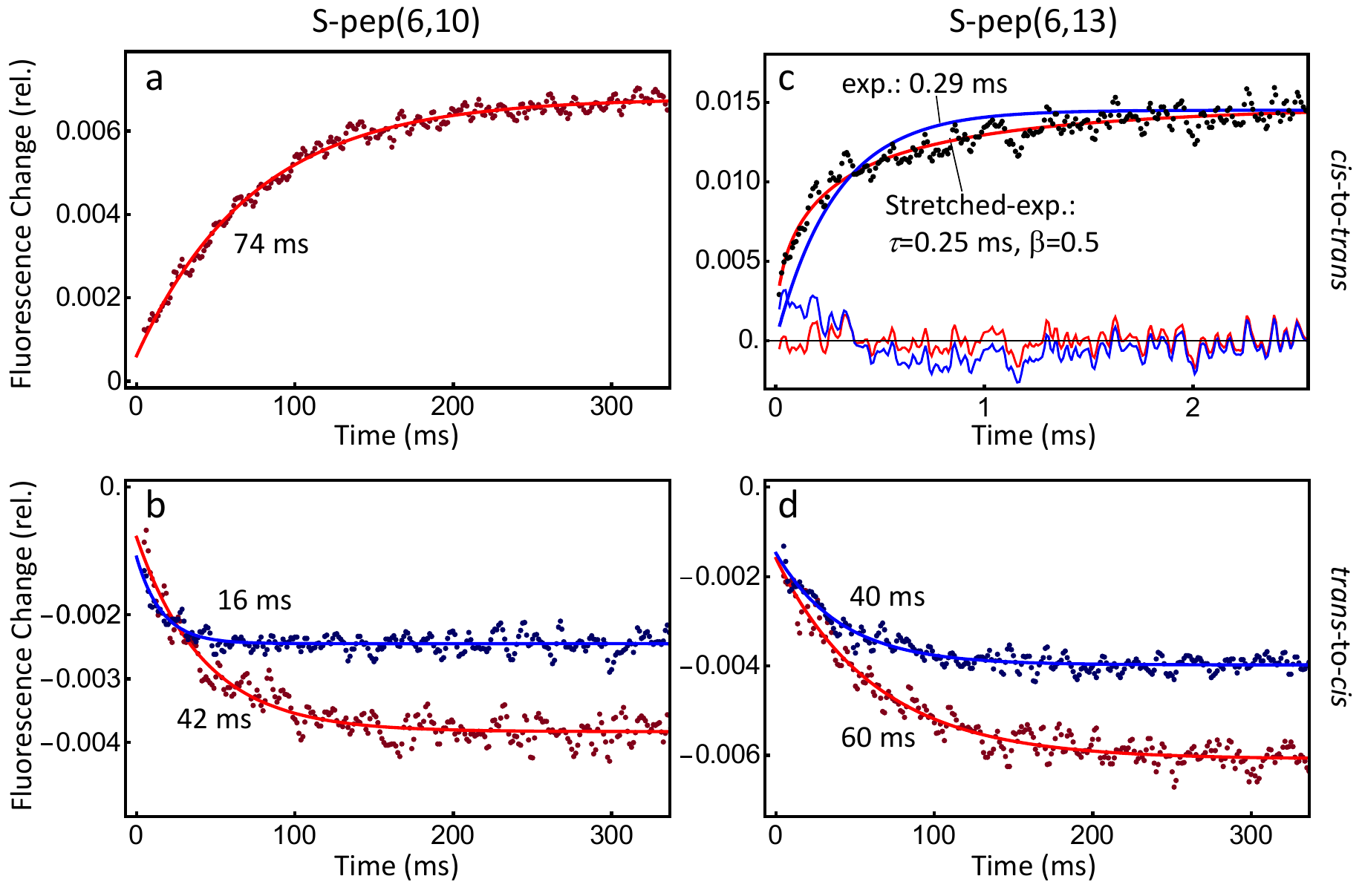}
		\caption{Time-resolved fluorescence. Relative fluorescence change for (a) \textit{cis}-to-\textit{trans} switching of S-pep(6,10) with 200~$\mu$M S-pep(6,10) and 200~$\mu$M S-protein, and (b) \textit{trans}-to-\textit{cis} switching with S-pep(6,10) concentration 200~$\mu$M, and S-protein concentration 100~$\mu$M (red) as well as 400~$\mu$M (blue). The same for (c) \textit{cis}-to-\textit{trans} switching of S-pep(6,13) with 400~$\mu$M S-pep(6,13) and 200~$\mu$M S-protein, and (d) \textit{trans}-to-\textit{cis} switching with S-pep(6,13) concentration 400~$\mu$M, and S-protein concentration 100~$\mu$M (red) as well as 400~$\mu$M (blue).The data in panels (a,b,d) are filtered with a Gaussian function with a width of 2.5~ms, those in panel (c) with with a width of 20~$\mu$s. In panels (a,b,d), the lines are exponential fits with the time constants indicated. In panel (c), the red line shows a stretched exponential fit ($\tau$=0.25~ms, stretching factor $\beta$=0.5), and the blue line a single exponential fit ($\tau$=0.29~ms). The residuals of these fits are shown as thin colored lines around the zero-line.} \label{FigExp}
	\end{center}
\end{figure*}

\subsection{S-pep(6,10)}

To set the stage, we start with  S-pep(6,10), whose binding affinities change by a large factor (20 fold) upon \textit{cis}-to-\textit{trans} switching, but with specific binding in both states (see Table~\ref{tab:Kd}). Fig.~\ref{FigExp}a shows the transient fluorescence measurement for \textit{cis}-to-\textit{trans} switching. Fluorescence increases, as expected since the tyrosines are quenched less upon unbinding of the S-peptide. An exponential fit to the data reveals a time-constant of 74~ms. Upon \textit{trans}-to-\textit{cis} switching (Fig.~\ref{FigExp}b), the sign of the signal inverts, representing stronger quenching of the tyrosine fluorescence upon ligand binding. The kinetics is concentration dependent with time constants of 42~ms for 100~$\mu$M S-protein (red) and 16~ms for 400~$\mu$M S-protein (blue, the concentration of the S-peptide has been 200~$\mu$M in both cases), as anticipated for a bimolecular reaction. The ratio of time constants closely resembles that of the S-protein concentrations. Furthermore, the amplitude of the 400~$\mu$M data (blue) is smaller, since we plot relative fluorescence change. In absolute numbers, the isomerized S-peptide is the same in both experiments (i.e., about 5\% of the S-peptide or 10~$\mu$M). In a relative sense, this is less for the large S-protein concentration.

\begin{figure*}[t]
	\centering
	\begin{center}
		\includegraphics[width=.8\textwidth]{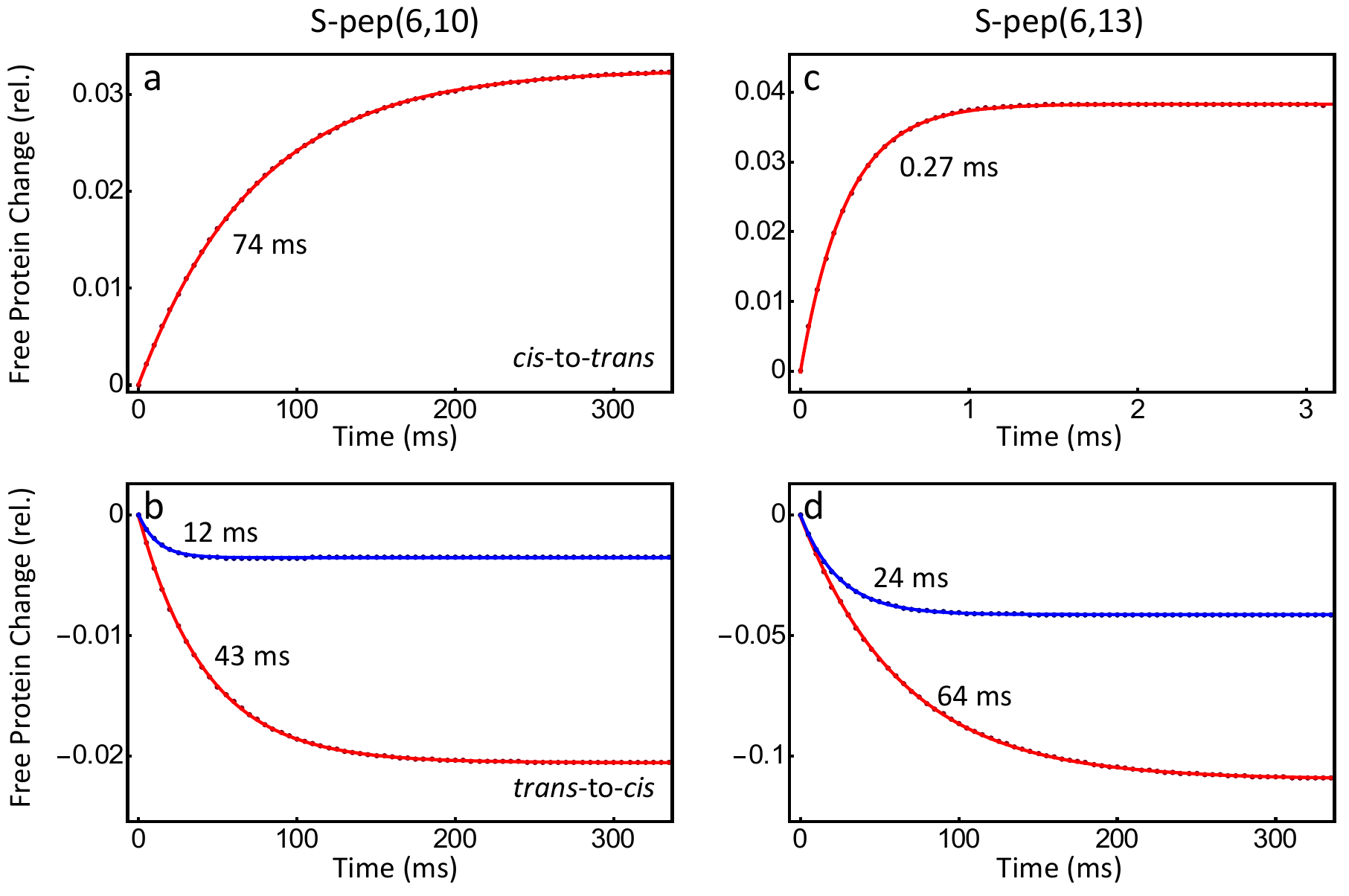}
		\caption{Model calculations.  Change of free S-protein upon (a) \textit{cis}-to-\textit{trans} and (b) \textit{trans}-to-\textit{cis} switching of S-pep(6,10), and upon (c) \textit{cis}-to-\textit{trans} and (d) \textit{trans}-to-\textit{cis} switching of S-pep(6,13), as deduced from the model described in Methods. The concentrations are the same as in the experiment (Fig.~\ref{FigExp}). The points are the result from the model, the solid lines exponential fits to it, with the fitted time constants indicated. }\label{FigSim}
	\end{center}
\end{figure*}

In its simplest form, ligand binding/unbinding is discussed in terms of the following chemical equilibrium:
\begin{equation}
PL \xrightleftharpoons[k_{on}]{k_{off}} P+L\label{eqEq}
\end{equation}
where $PL$ is the ligand-bound state, and $P$ and $L$ denote protein and ligand, respectively.
The dissociation constant $K_d$ is related to the rate constants $k_{on}$ and $k_{off}$  by:
\begin{equation}
K_d=\frac{k_{off}}{k_{on}}. \label{eqKd}
\end{equation}
The equilibrium experiments of Ref.~\onlinecite{jankovic2019} can only determine the dissociation constant $K_d$, while the present kinetic experiments can also determine the on- and off-rate constants. Trends can be read off directly from Fig.~\ref{FigExp}a,b, but a more quantitative modelling is needed to extract these rate constants, taking into account the fact that both states of the photoswitch bind to the protein to a certain extent, and competitive binding of the S-peptide in its \textit{cis} and \textit{trans}-states, both of which exist after photoswitching (see Fig.~\ref{FigSim}, for details of that model, see Methods). The resulting kinetics are not strictly exponential owing to the coupled and nonlinear character of the corresponding differential equations, however, they deviate from exponential by less than what the experimental noise would allow one to see (Fig.~\ref{FigSim}). We therefore also fit the simulated data to exponential functions, and compare the extracted time-constants with the experimental ones. We obtain good qualitative agreement for both \textit{cis}-to-\textit{trans} and \textit{trans}-to-\textit{cis} switching when assuming on-rate constants  $k_{on,cis}=3\cdot10^5$~M$^{-1}$s$^{-1}$ and $k_{on,trans}=1\cdot10^5$~M$^{-1}$s$^{-1}$, see Table~\ref{tab:Kd}.

\begin{table}[b]
\centering
\caption{Thermodynamic and kinetic constants for the two samples considered in this study.}
\label{tab:Kd}
\begin{tabular}{l |  c| c |c |c }
            & $K_{d,cis}$                & $k_{on,cis}$                     & $K_{d,trans}$             & $k_{on,trans}$         \\\hline
S-pep(6,10) &2.3~$\mu$M\footnotemark[1]  & $3\cdot10^5$~M$^{-1}$s$^{-1}$    & 47~$\mu$M\footnotemark[1] & $1\cdot10^5$~M$^{-1}$s$^{-1}$  \\
S-pep(6,13) &70~$\mu$M\footnotemark[1]   & $9\cdot10^4$~M$^{-1}$s$^{-1}$    & 40~mM\footnotemark[2]	                    & $9\cdot10^4$~M$^{-1}$s$^{-1}$ \footnotemark[3] \\
\end{tabular}
\footnotetext[1]{taken from Ref.~\onlinecite{jankovic2019}}
\footnotetext[2]{nominal dissociation constant; see text for discussion.}
\footnotetext[3]{not measured, but assumed to be the same as $k_{on,cis}$, see text for discussion}
\end{table}

The extracted on-rate constants are in the same range as what has been observed in Refs.~\onlinecite{goldberg1999,bachmann2011} for a series of mutants of the RNase~S system without photoswitch. Two factors determine binding rate constants. The first is the diffusion controlled formation of an encounter complex, taking into account the fact that the two partners need to approach each other with a specific orientation, which results in typical on-rate constants in the range between $10^5$~M$^{-1}$s$^{-1}$ to $10^6$~M$^{-1}$s$^{-1}$.\cite{schreiber2009,rogers2013} The second factor concerns the fraction of molecules that leave the encounter complex before a stable protein-ligand complex is formed (in an induced fit scenario). Since the diffusion controlled step is often rate-limiting, $k_{on}$ typically varies only in a small range, in the case of the mutants of the RNase~S system between $1.6\cdot10^5$~M$^{-1}$s$^{-1}$ and $5.8\cdot10^5$~M$^{-1}$s$^{-1}$, see Ref.~\onlinecite{bachmann2011}. In our case, the \textit{cis}-state is more tightly bound, and correspondingly, $k_{on,cis}$ is $\approx$3 times faster than $k_{on,trans}$, however, this factor 3 is small in comparison to the overall factor 20, by which the binding affinities differ.

\subsection{S-pep(6,13)}

With that, we turn to S-pep(6,13), which is characterized as on-off system without any specific binding detected in the \textit{trans}-state.\cite{jankovic2019} At a first sight, the results in Fig.~\ref{FigExp}c,d look similar to those of S-pep(6,10) (Fig.~\ref{FigExp}a,b), however, binding and unbinding happens on completely different timescales (to that end note the different time ranges in Figs.~\ref{FigExp}c,d). Binding upon \textit{trans}-to-\textit{cis} switching, which is again concentration dependent, reveals 60~ms for 100~$\mu$M S-protein concentration and 40~ms for 400~$\mu$M S-protein concentration, with the S-peptide concentration being 400~$\mu$M in both cases (Fig.~\ref{FigExp}d). These are similar timescales as for S-pep(6,10). However, the ratio of time constants deviates significantly from that of the S-protein concentrations, for two reasons: First, the observed rate constant is the sum of the effective (i.e., concentration dependent) on- and off-rate constants, and the off-rate contributes more in a relative sense at lower protein concentrations. This effect is taken care of in the model of Fig.~\ref{FigSim}d, where a ratio of time constant smaller than 4 is indeed observed. In addition, by comparing CD with fluorescence quenching data, we concluded in Ref.~\onlinecite{jankovic2019} that some amount of non-specific binding also occurs in the \textit{trans}-state of S-pep(6,13). Binding of this fraction of molecules will be a unimolecular process, and thus not concentration dependent.

Unbinding upon \textit{cis}-to-\textit{trans} switching is faster by two orders of magnitudes, see Fig.~\ref{FigExp}c. Furthermore, the data reveal stretched-exponential kinetics $\exp[-(t/\tau)^\beta]$ with time-constant $\tau$=0.25~ms and a significant stretching factor $\beta$=0.5. From the residual shown in Fig.~\ref{FigExp}c, it is evident that the stretched exponential fit is better than a single-exponential fit with time-constant 0.29~ms (Fig.~\ref{FigExp}c, blue line), which we consider an average time-constant.

The binding affinity of the \textit{trans}-state could not be measured in Ref.~\onlinecite{jankovic2019}, since it is too small. For a modelling similar to the one used for S-pep(6,10), we therefore had to assume $k_{on,cis}=k_{on,trans}$, revealing $k_{on}\approx9\cdot10^4~\rm{M}^{-1}\rm{s}^{-1}$ and a nominal dissociation constant of $K_d=40$~mM,  see Figs.~\ref{FigSim}c,d. The modelled unbinding kinetics of Fig.~\ref{FigSim}c is almost perfectly exponential, in contrast to the experimental data of Fig.~\ref{FigExp}c. Since binding essentially does not exist, $k_{on}$ is negligible, and the reaction is unimolecular to a very good approximation; very different from typical binding studies.\cite{goldberg1999,bachmann2011} This is due to the very special molecular system, which has been designed to not bind upon switching into the \textit{trans}-state. The stretched-exponential character of the experimental data thus must be due to an effect that is beyond Eq.~\ref{eqEqcouled}.

\section{Discussion and Conclusion}

A dissociation constant of $K_d=40$~mM  is no longer a meaningful number; for example, with the molecular weight of protein plus peptide (14 kDa), one would conclude that a 50\% binding equilibrium is reached only at a protein density of 560~g/l, i.e., in a sample that contains about the same amount of water as it contains protein. This density is beyond a regime, in which Eq.~\ref{eqEq} is valid. In other words, the peptide does not bind at all to the protein in the \textit{trans}-state and unbinding is a barrier-less process. It is important to stress that this conclusion originates from the averaged time-constant of unbinding (0.29~ms), irrespective whether the kinetics is exponential or stretched-exponential. This time-constant is two to five orders of magnitudes faster than that of the mutants of the RNase~S system studied in Ref.~\onlinecite{bachmann2011}. One assumption had to be made to estimate the binding affinity in the \textit{trans}-state, namely that $k_{on,trans}=k_{on,cis}$, since only $k_{on,cis}$ is determined by the experimental data of Fig.~\ref{FigExp}d. This assumption is justified by the fact that on-rate constants vary only by relatively small amounts. Taking  Ref.~\onlinecite{bachmann2011} as a basis with a variation of $k_{on}$ of a factor 3-4 for a wide variety of S-peptide mutants, this gives a feeling of the uncertainty in the estimate of $K_d=40$~mM. Even considering this uncertainty, one would conclude that the ligand in essence does not bind in the \textit{trans}-state of the photoswitch.

\begin{figure}[t]
	\centering
	\begin{center}
		\includegraphics[width=.5\textwidth]{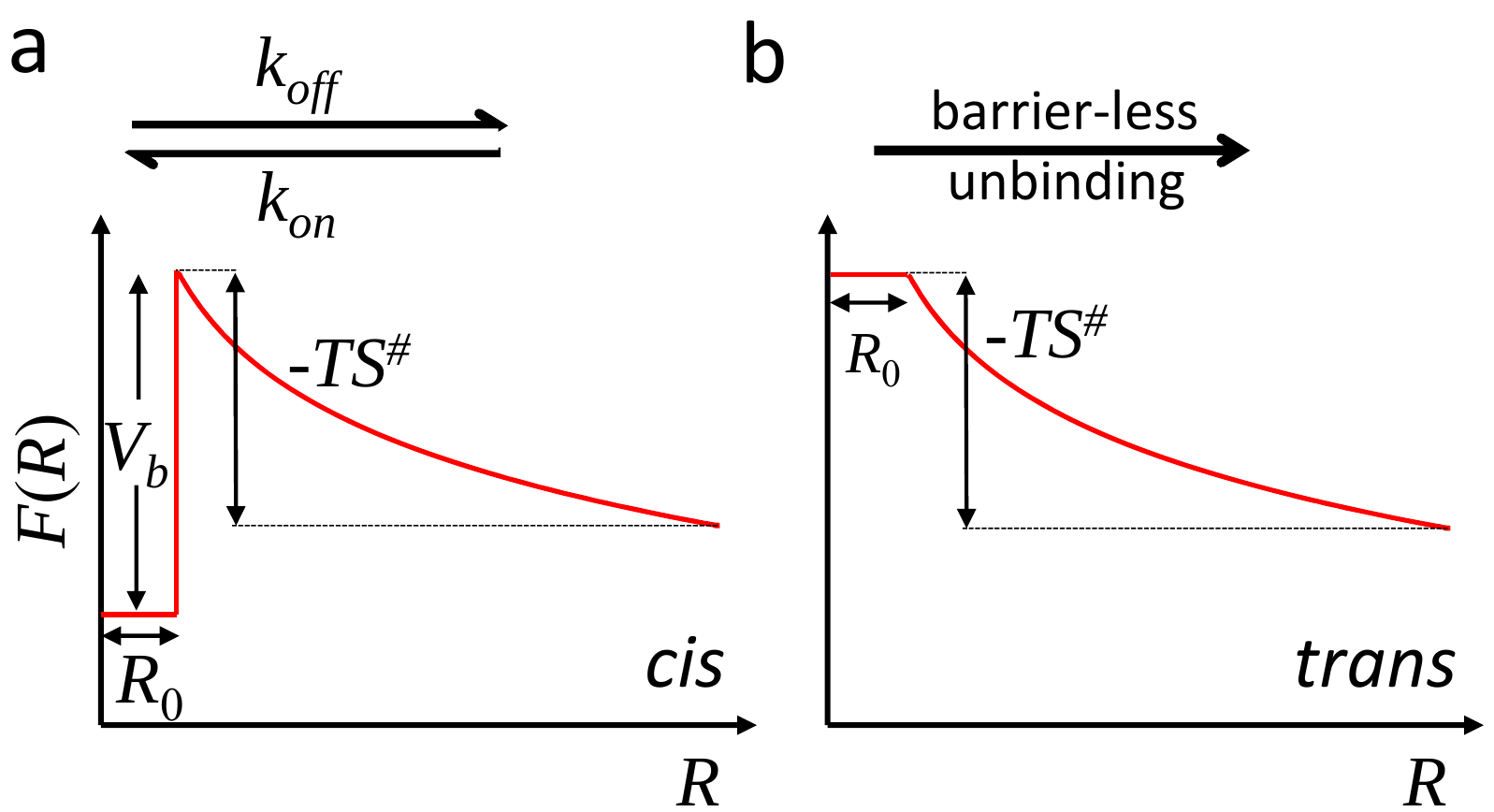}
		\caption{Free energy model: (a) Free energy of the ligand as a function of the distance $R$ of the ligand from the protein, where $R_0$ is the interaction range of the protein (in essence its size), $V_b$ the binding energy, and $S^\#$ an activation entropy. Panel (b) shows the same for barrier-less unbinding with $V_b=0$. The two scenarios resemble the situations in the \textit{cis}- and \textit{trans}-states of the photoswitch.}\label{FigFreeEnergy}
	\end{center}
\end{figure}

To illustrate the concept of barrier-less unbinding, Fig.~\ref{FigFreeEnergy}a shows a very simple model for the free energy of ligand binding as a function of the distance $R$ of the ligand from the protein.\cite{bicout00} The protein-ligand complex is stabilized by a binding energy $V_b$. Beyond the interaction range of the protein ($R_0$), the free energy decreases due to an entropic contribution,
which accounts for the larger space available to the ligand with increasing distance. The binding energy $V_b$ determines an energetic barrier for unbinding, while the barrier for binding is in essence of entropic nature. When the binding energy becomes zero, $V_b=0$, unbinding is barrier-less, see Fig.~\ref{FigFreeEnergy}b. Upon \textit{cis}-to-\textit{trans}-isomerisation, we switch from the free energy of Fig.~\ref{FigFreeEnergy}a to that of Fig.~\ref{FigFreeEnergy}b. The ensemble, which has been equilibrated on the free energy surface Fig.~\ref{FigFreeEnergy}a, all of the sudden finds itself in a non-equilbrium situation, and starts to evolve on the free energy surface Fig.~\ref{FigFreeEnergy}b.

In the realm of femtochemistry,\cite{zewail00} it is important to distinguish ``kinetics'' from ``dynamics''. The on- and off-rate constants in Eq.~\ref{eqEq} determine the probability of ligand binding and unbinding per time unit, and completely mask the complexity of the process. This approach is valid when the barriers are high enough so that their crossing becomes rate-limiting. In that limit, one can describe the kinetics by single numbers, $k_{on}$ and $k_{off}$. On the other hand, it is clear from MD simulations that ligand binding or unbinding, when looked at on an atomistic level, is a very complex and very heterogenous process, with different pathways consisting of many small steps.\cite{Blochliger2015,Luitz2017,robustelli2020} That is what we call the ``intrinsic dynamics'' of the process. When removing the unfolding barrier in S-pep(6,13), we take a glance at the intrinsic dynamics. The stretched-exponential function is a commonly chosen model to account for a distribution of timescales, with the stretching factor $\beta$ determining the width of that distribution.\cite{Johnston2006} In the present case, a distribution of timescales could result from the final unbinding of the S-peptide from various unspecific binding sites on the protein surface. Molecular dynamics (MD) simulations would help to provide more microscopic insights into the various pathways that give rise to the distribution of timescales we observe.


The situation strongly resembles that of ``downhill'' protein folding. When the folding barrier of a protein is removed, e.g., by mutations, one observes the intrinsic dynamics of the protein, sometimes also called the ``speed limit'' of protein folding.\cite{Yang03,kubelka04} At the same time, the dynamics becomes pronouncedly non-exponential,\cite{sabelko99} just like barrier-less unbinding in Fig.~\ref{FigExp}c. Similar to the protein folding problem,\cite{Ma2005} it is expected that other probes, e.g., transient IR spectroscopy (which is currently ongoing in our lab), will reveal different dynamical components of the process. As a word of caution, it should however be added that the implication of downhill folding for non-exponential kinetics has been questioned.\cite{hagen03}

One of the scenarios discussed in the context of ligand binding is that of an induced fit, which is described by the following reaction scheme:
\begin{equation}
P+L \xrightleftharpoons[k_{off}]{k_{on}} PL^\# \xrightleftharpoons[k'_{r}]{k_{r}} PL \label{eqEqind}
\end{equation}
Here, $PL^\#$ is a high-energy bound state that relaxes into $PL$ upon ``fitting'' the ligand into the binding site of a protein. Since the binding rate constant is concentration dependent, the diffusive step will not be rate-limiting at high enough concentrations, while the second step remains concentration independent. Observing an effective rate constant of binding that saturates with increasing concentration is considered an indicator for an induced fit.\cite{vogt2012, gianni2014, paul2016} However, if the rate constant $k_r$ of the induced fit is too fast, this approach might miss it, as it is not possible to increase the concentrations sufficiently. For typical on-rate constants of $10^5-10^6$~M$^{-1}$s$^{-1}$,\cite{schreiber2009} and typical maximal protein concentrations of 1~mM, that regime is already reached for $k_r>10^2-10^3$~s$^{-1}$. In addition to this inherent limitation, the time resolution of typical stop-flow instruments is in the range of 1~ms.\cite{gianni2014}

The experiment we perform here is closely related, as it can also be described by Eq.~\ref{eqEqind} (where $PL^\#$ would be the transition state in Fig.~\ref{FigFreeEnergy}), just that we consider here the unfolding direction. The forward and backward rate constants $k_r$ and $k'_r$ are connected to each other by the equilibrium constant of the second step. In the barrier-less case of Fig.~\ref{FigFreeEnergy}b, the equilibrium constant is 1, and $k_r=k'_r$. We observe $k'_r=4\cdot10^3$~s$^{-1}$, which is fast in light of the discussion above, but slow in terms of the structural rearrangements that are needed to fit the S-peptide into the binding site of the S-protein. That is, the folding of small $\alpha$-helices in solution occurs on a typically 3-4 orders of magnitude faster timescale ($10^6-10^7$~s$^{-1}$),\cite{kubelka04} even under constraints.\cite{Ihalainen2008}

In conclusion, due to the slow diffusive step inherent to any binding experiments, many induced fit scenarios might be missed in such experiments.  In connection with a fast trigger, much quicker structural processes can be observed in the unbinding direction, revealing the intrinsic dynamics of the ligand during the unbinding event. That ``speed limit'' is in the range of few 100~$\mu$s for the RNase~S system.\\


\noindent\textbf{Acknowledgement:} We thank Claudio Zanobini and Karl Hamm for technical contributions at an early stage of this project. We also thank Rolf Pfister for the synthesis of the photoswitch (BSBCA), the Functional Genomics Center
Zurich, especially Serge Chesnov, for his work on the mass spectrometry. The work has been supported by the Swiss National Science Foundation (SNF) through the NCCR MUST and Grant 200020B\_188694/1.\\


\noindent\textbf{Accession Codes:} Ribonuclease A (P61823 (RNAS1\_BOVIN)) and Subtilisin (P00782 (SUBT\_BACAM))\\

%

\makeatletter
\def\@biblabel#1{(#1)}
\makeatother

\def\bibsection{\section*{}} 

\noindent\textbf{References:}
\vspace{-1.5cm}


\providecommand{\latin}[1]{#1}
\makeatletter
\providecommand{\doi}
  {\begingroup\let\do\@makeother\dospecials
  \catcode`\{=1 \catcode`\}=2 \doi@aux}
\providecommand{\doi@aux}[1]{\endgroup\texttt{#1}}
\makeatother
\providecommand*\mcitethebibliography{\thebibliography}
\csname @ifundefined\endcsname{endmcitethebibliography}
  {\let\endmcitethebibliography\endthebibliography}{}

%
%

\end{document}